\documentclass[aps,prd,reprint,twocolumn,superscriptaddress,showpacs]{revtex4-1}
\usepackage{graphicx}
\usepackage{mathrsfs}
\usepackage{bm}
\usepackage{amsmath}
\usepackage{dcolumn}
\usepackage{epstopdf}
\usepackage{dsfont}
\usepackage{amssymb}
\usepackage{tabularx}
\usepackage{array}
\usepackage{float}
\usepackage{color}
\usepackage{epstopdf}
\usepackage{mathrsfs}
\usepackage[colorlinks, linkcolor=blue,anchorcolor=blue,citecolor=blue,urlcolor=blue]{hyperref}
\usepackage{multirow}
\usepackage{booktabs}

\begin{document}
\title{Hexagonal supertetrahedral boron: A topological metal with multiple spin-orbit-free emergent fermions}

\author{Yan Gao}
\email{These two authors contributed equally to this work.}
\affiliation{Department of physics and Beijing Key Laboratory of Opto-electronic Functional Materials $\&$ Micro-nano Devices, Renmin University of China, Beijing 100872, China}

\author{Weikang Wu}
\email{These two authors contributed equally to this work.}
\affiliation{Research Laboratory for Quantum Materials, Singapore University of Technology and Design, Singapore 487372, Singapore}

\author{Peng-Jie Guo}
\affiliation{Department of physics and Beijing Key Laboratory of Opto-electronic Functional Materials $\&$ Micro-nano Devices, Renmin University of China, Beijing 100872, China}

\author{Chengyong Zhong}
\affiliation{Institute for Advanced Study, Chengdu University, Chengdu 610106, China}

\author{Shengyuan A. Yang}
\email{shengyuan\_yang@sutd.edu.sg}
\affiliation{Research Laboratory for Quantum Materials, Singapore University of Technology and Design, Singapore 487372, Singapore}
\affiliation{Center for Quantum Transport and Thermal Energy Science, School of Physics and Technology, Nanjing Normal University, Nanjing 210023, China}

\author{Kai Liu}
\email{kliu@ruc.edu.cn}
\affiliation{Department of physics and Beijing Key Laboratory of Opto-electronic Functional Materials $\&$ Micro-nano Devices, Renmin University of China, Beijing 100872, China}

\author{Zhong-Yi Lu}
\email{zlu@ruc.edu.cn}
\affiliation{Department of physics and Beijing Key Laboratory of Opto-electronic Functional Materials $\&$ Micro-nano Devices, Renmin University of China, Beijing 100872, China}

\begin{abstract}
We predict a new three-dimensional (3D) boron allotrope based on systematic first-principles electronic structure calculations. This allotrope can be derived by substituting each carbon atom in a hexagonal diamond lattice with a B$_{4}$ tetrahedron and possesses the same space group $P6_{3}/mmc$ as hexagonal diamond, hence it is termed as H-boron. We show that H-boron has good stability and excellent mechanical property. Remarkably, we find that H-boron is a topological metal with rich types of spin-orbit-free emergent fermions, including semi-Dirac fermion, quadratic and linear triple-point fermion, nodal-line fermion, and nodal-surface fermion. We clarify their symmetry protections and characterize them by constructing the corresponding low-energy effective models. Our work not only discovers a new boron allotrope with excellent properties, it also offers a platform to explore interesting physics of new kinds of emergent fermions.
\end{abstract}
\maketitle

\section{Introduction}\label{sec_introduction}

Boron is the left neighbor of carbon in the periodic table. Like carbon, boron can form a diversity of allotropic structures, including zero-dimensional (0D) clusters, 1D nanotubes, 2D sheets, and also 3D polymorphs~\cite{1Zhang2017,2Sun2017,3Reddy2016}. In recent years, along with the rise of 2D materials, several 2D boron structures have been predicted~\cite{4Xiaobao2008,5Xiaojun2012,6Tang2007,7Sohrab2010,8Ozdogan2010}, among which honeycomb borophene~\cite{9Weifeng2018}, triangular borophene~\cite{10Mannix2015}, $\beta_{12}$ and $\chi_{3}$ boron sheets~\cite{11Feng2016} have already been successfully synthesized in experiments. The 3D boron allotropes are also quite abundant~\cite{12Ogitsu2013}, such as the experimentally prepared $\alpha$-Ga-type boron~\cite{13Haussermann2003}, $\gamma$-B$_{28}$~\cite{14Gatti2009} and $\alpha$-B$_{12}$~\cite{15Solozhenko2009}. In comparison to carbon, boron is short of one electron, which results in large difference in their bonding properties and hence structural features. For example, the cubic and hexagonal diamond structures become unstable if the carbon atoms are each replaced by boron. Particularly, it is noticed that quite a number of 3D stable boron allotropes comprise the icosahedral B$_{12}$ clusters~\cite{16Hillebrecht2009}, which appear as basic building blocks for 3D boron structures. It is interesting to ask whether or not there exist new 3D boron allotropes consisting of other kinds of motifs.

Meanwhile, the study of topological materials has attracted great interest in recent physics and materials research. In topological metals (including semimetals), novel quasiparticles emerge around the protected band crossing point, some of which are beyond the elementary fermions in high-energy physics and possess fascinating physical properties~\cite{17Binghai2017,18Armitage2018,19Richard2017,20Bansil2016,21Shengyuan2016}. While most works in this field have been focused on materials involving heavy elements, expecting that nontrivial band topology may be driven by the strong spin-orbit coupling (SOC), it is also realized that light-element materials offer a distinct alternative. For materials made of light elements such as boron or carbon, SOC effect can be neglected. Therefore, the electron spin can be regarded as a dummy degree of freedom, and the emergent fermions can be regarded as ``spinless" or ``spin-orbit-free", which are \emph{fundamentally distinct} from the ``spinful" fermions in materials with sizable SOC~\cite{22Schnyder2008}. Previous works on carbon have revealed several 3D allotropes as topological metals, hosting Weyl fermions~\cite{23Cohen2015}, nodal-loop fermions~\cite{23Cohen2015,24Hongming2016,25Yuanping2018,26Sung2017,27Kawazoe2015,28Mullen2015}, triple-point fermions~\cite{29Vanderbilt2018,30Zhongcy2017}, Hopf-link fermions~\cite{30Zhongcy2017}, and even nodal-surface fermions~\cite{31ZhongCohen2016,32Weikang2018}. Due to their distinct structural features, one may naturally expect that 3D boron allotropes may be a good platform to search for new kinds of spin-orbit-free fermions. Currently, the exploration of this research direction has just begun~\cite{33GaoYan2018,34Yazyev2018}.

\begin{figure*}[!t]
	\centering
	\includegraphics[width=0.7\textwidth]{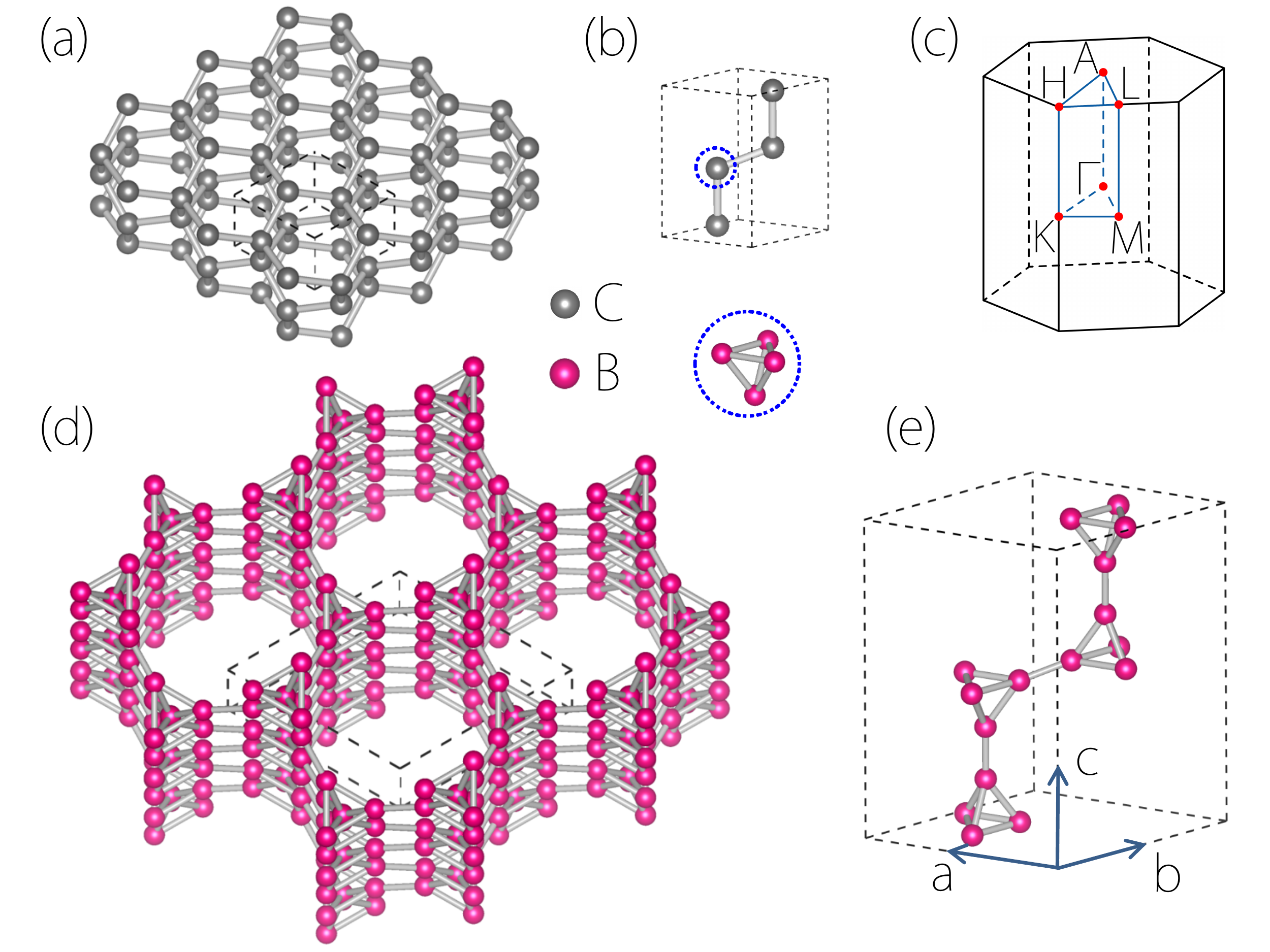}
	\caption{(a) Structure of a hexagonal diamond (h-diamond). (b) Primitive cell of h-diamond. (c) Brillouin zone (BZ) of H-boron. (d) Side view of the optimized structure of hexagonal supertetrahedral boron (H-boron), which can be derived by the substitution of each carbon atom in h-diamond by a B$_4$ tetrahedron. (e) Primitive cell of H-boron.}
	\label{fig_structure}
\end{figure*}

In this work, we predict a new 3D boron allotrope which is termed as H-boron. The prediction is inspired by the previous work of Sheng and coworkers on proposing T-carbon~\cite{35ShengXL2011}, in which each site of a cubic diamond (c-diamond) lattice is occupied by a carbon tetrahedron. Later, T-carbon nanowires were successfully realized in experiment via laser irradiation method~\cite{36Jinying2017}. A boron allotrope called cF-B$_{8}$ with the same structure as T-carbon was also predicated recently~\cite{37Getmanskii2017}, indicating that cubic diamond structure combined with B$_{4}$ tetrahedron motif may stabilize a 3D boron structure. Given that hexagonal diamond (h-diamond) shares similar stability as c-diamond, it is natural to expect that B$_{4}$ tetrahedrons arranged into an h-diamond lattice may also generate a stable 3D boron allotrope. We find that this is indeed the case and the resulting material is the H-boron. By means of systematic density functional theory (DFT) calculations, we study the structure, stability, and electronic properties of H-boron. We show that H-boron is both dynamically and thermally stable. The \emph{ab initio} molecular dynamics simulation indicates that H-boron can be stable up to 600~K. Notably, the material has a low mass density of $\sim$0.91~g/cm$^{3}$, even less than water, while its bulk modulus is larger than glass, indicating its potential application as aerospace and cosmic materials. Most interestingly, we find that H-boron is a new topological metal hosting multiple types of spin-orbit-free emergent fermions, including semi-Dirac fermion, quadratic and linear triple-point fermion, nodal-line fermion, and nodal-surface fermion. Particularly, the 3D semi-Dirac fermion and the quadratic triple-point fermion have not been discussed before. We also suggest a possible scheme for synthesizing H-boron in experiment.

\section{Computational details}\label{sec_computation}
The first-principles electronic structure calculations were based on the density functional theory (DFT) with the projector augmented wave (PAW) method~\cite{38Blöchl1994} as implemented in the VASP package~\cite{39Kresse1996}. The generalized gradient approximation (GGA) of the Perdew-Burke-Ernzernof (PBE) type~\cite{40Perdew1996} was adopted for the exchange-correlation functional. The kinetic energy cutoff of the plane wave basis was set to 550~eV. A $9 \times 9 \times 7$ Monkhorst-Pack $k$-point mesh~\cite{41Monkhorst1976} was taken for the Brillouin zone (BZ) sampling, and the Gaussian smearing method with a width of 0.05~eV was used. Both the lattice parameters and the internal atomic positions were fully relaxed until the forces on all the atoms were less than 0.001~eV/\text{\AA}. To study the dynamical stability, we performed phonon calculations using the Phonopy package~\cite{42Togo2008}, for which a $2 \times 2 \times 2$ supercell containing 128 atoms was used. The thermal stability was investigated with the \emph{ab initio} molecular dynamics (AIMD) simulations in a canonical ensemble~\cite{43NoseS1984} with a Nos\'{e}-Hoover thermostat. The band crossing features were studied by using the WannierTools~\cite{44ShengNan2018} package based on the tight-binding model constructed via the Wannier90 code~\cite{45Mostofi2008}.

\section{Results and analysis}\label{sec_results}

\begin{figure}[t]
	\centering
	\includegraphics[width=0.48\textwidth]{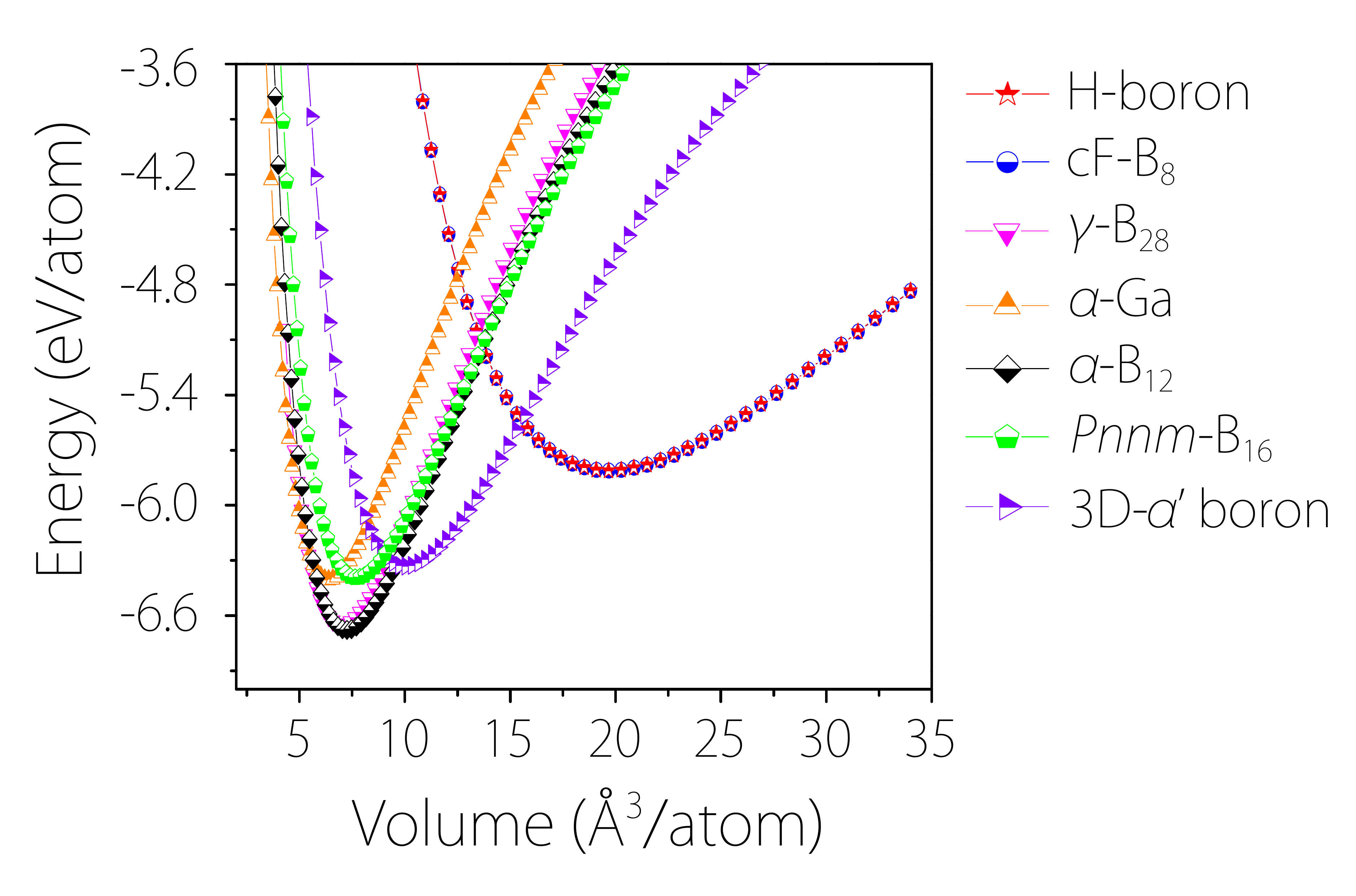}
	\caption{Total energy per atom as a function of volume (equation of states, EOS) for typical boron allotropes, including H-boron, cF-B$_{8}$, $\gamma$-B$_{28}$, $\alpha$-Ga-type boron, $\alpha$-B$_{12}$, $Pnnm$-B$_{16}$, and 3D-$\alpha'$ boron allotropes.}
	\label{fig_energy}
\end{figure}

{\renewcommand{\arraystretch}{1.35}
\begin{table*}[th]
\caption{\label{tab:I} Space groups, lattice constants (\text{\AA}), angles ($^{\circ}$), Wyckoff positions, densities (g/cm$^{3}$), bond lengths (\text{\AA}), bulk moduluses (GPa), and cohesive energies E$_{coh}$ (eV/B) for H-boron, cF-B$_{8}$, $\alpha$-Ga, $\gamma$-B$_{28}$, and $\alpha$-B$_{12}$, respectively.}
\resizebox{\textwidth}{25mm}{
\begin{tabular}{|c|c|c|c|c|c|c|c|c|c|c|c|c|c|c|}
\hline
\multicolumn{2}{|c|}{\multirow{2}{*}{Structures}} & \multirow{2}{*}{Space groups} & \multicolumn{3}{c|}{Lattice parameters (\text{\AA})} & \multicolumn{2}{c|}{Angles($^{\circ}$)} & \multicolumn{3}{c|}{Wyckoff positions} & \multirow{2}{*}{Densities (g/cm$^{3}$)} & \multirow{2}{*}{Bond lengths (\text{\AA})} & \multirow{2}{*}{Bulk moduluses (GPa)} & \multirow{2}{*}{E$_{coh}$ (eV/B)} \\

\cline{4-11}
\multicolumn{2}{|c|}{} &  & $a$ & $b$ & $c$ & $\alpha$=$\beta$ & $\gamma$ & $x$ & $y$ & $z$ &  &  &  &  \\

\hline

\multirow{2}{*}{H-boron}         & B1($12k$)      & \multirow{2}{*}{$P6_{3}/mmc$}      & \multirow{2}{*}{6.06} & \multirow{2}{*}{6.06} & \multirow{2}{*}{9.91} & \multirow{2}{*}{90}    & \multirow{2}{*}{120}   & 0.5727      & 0.4273     & 0.5277     & \multirow{2}{*}{0.91}             & \multirow{2}{*}{1.62-1.71}       & \multirow{2}{*}{70.42}               & \multirow{2}{*}{-5.81}      \\

\cline{2-2}\cline{9-11}
  & B2($4f$) &  &  &  &  &  &  & 0.3333 & 0.6667 & 0.3319 &  &  &  &  \\

\hline

cF-B$_{8}$ & B1($32e$) & $Fd\bar{3}m$ & 8.57 & 8.57 & 8.57 & 90 & 90 & -0.1796 & 0.3204 & 0.1796 & 0.91 & 1.62-1.71 & 70.42 & -5.81  \\

\hline

$\alpha$-Ga & B1($8f$) & $Cmca$ & 2.94 & 5.33 & 3.26 & 90 & 90 & 0 & 0.1558 & 0.0899 & 2.81 & 1.76-1.92 & 264.30 & -6.41                       \\

\hline

\multirow{5}{*}{$\gamma$-B$_{28}$}       & B1($4g$)       & \multirow{5}{*}{$Pnnm$}         & \multirow{5}{*}{5.04} & \multirow{5}{*}{5.61} & \multirow{5}{*}{6.92} & \multirow{5}{*}{90}    & \multirow{5}{*}{90}    & 0.1702      & 0.5206     & 0          & \multirow{5}{*}{2.57}             & \multirow{5}{*}{1.66-1.90}       & \multirow{5}{*}{244.29}              & \multirow{5}{*}{-6.65}      \\

\cline{2-2}\cline{9-11}

  & B2($8h$) &  &  &  &  &  &  & 0.1606 & 0.2810 & 0.3743 &  &  &  &  \\

\cline{2-2}\cline{9-11}

  & B3($8h$) &  &  &  &  &  &  & 0.3472 & 0.0924 & 0.2093 &  &  &  &  \\

\cline{2-2}\cline{9-11}

  & B4($4g$) &  &  &  &  &  &  & 0.3520 & 0.2711 & 0 &  &  &  &  \\

\cline{2-2}\cline{9-11}

  & B5($4g$) &  &  &  &  &  &  & 0.1644 & 0.0080 & 0 &  &  &  &   \\

\hline
\multirow{2}{*}{$\alpha$-B$_{12}$}       & B1($18h$)      & \multirow{2}{*}{$R\bar{3}m$}          & \multirow{2}{*}{5.05} & \multirow{2}{*}{5.05} & \multirow{2}{*}{5.05} & \multirow{2}{*}{58.04} & \multirow{2}{*}{58.04} & 0.0103      & 0.0103     & 0.6540     & \multirow{2}{*}{2.48}             & \multirow{2}{*}{1.67-1.80}       & \multirow{2}{*}{237.16}              & \multirow{2}{*}{-6.68}      \\

\cline{2-2}\cline{9-11}

  & B2($18h$)  &  &  &  &  &  &  & 0.2211  & 0.2211  & 0.6305 &  &  &  & \\
\hline
\end{tabular}}
\end{table*}
}

The predicted H-boron was inspired by the proposal of T-carbon~\cite{35ShengXL2011}, which is constructed through replacing each carbon atom in c-diamond lattice by a C$_{4}$ tetrahedron. Instead of c-diamond lattice, we consider its close cousin, the h-diamond lattice, as illustrated in Fig.~\ref{fig_structure}(a). It is known that as carbon allotropes, hexagonal and cubic diamonds share very similar properties. In both structures, the atoms are bonded tetrahedrally. Here, we put a B$_{4}$ tetrahedron at each lattice site of an h-diamond lattice [Fig.~\ref{fig_structure}(b)], forming the boron structure shown in Fig.~\ref{fig_structure}(d). It possesses the same space group $P6_{3}/mmc$ as the original h-diamond lattice. The primitive cell of H-boron contains four tetrahedrons with sixteen boron atoms [Fig.~\ref{fig_structure}(e)]. The optimized lattice parameters are $a = b = 6.06~\text{\AA}$ and $c = 9.91~\text{\AA}$. The boron atoms occupy the Wyckoff positions $12k$ (0.5727, 0.4273, 0.5277) and $4f$ (0.3333, 0.6667, 0.3319), respectively. There exist two nonequivalent B-B bonds with respective lengths of 1.62~\text{\AA} (inter-tetrahedron) and 1.71~\text{\AA} (intra-tetrahedron), which are comparable to those in the experimentally synthesized 3D boron allotropes~\cite{13Haussermann2003,14Gatti2009,15Solozhenko2009}, such as in $\alpha$-B$_{12}$ (1.67-1.80~\text{\AA}), $\gamma$-B$_{28}$ (1.66-1.90~\text{\AA}), and $\alpha$-Ga-type boron (1.76-1.92~\text{\AA}).

\begin{figure}[t]
	\centering
	\includegraphics[width=0.48\textwidth]{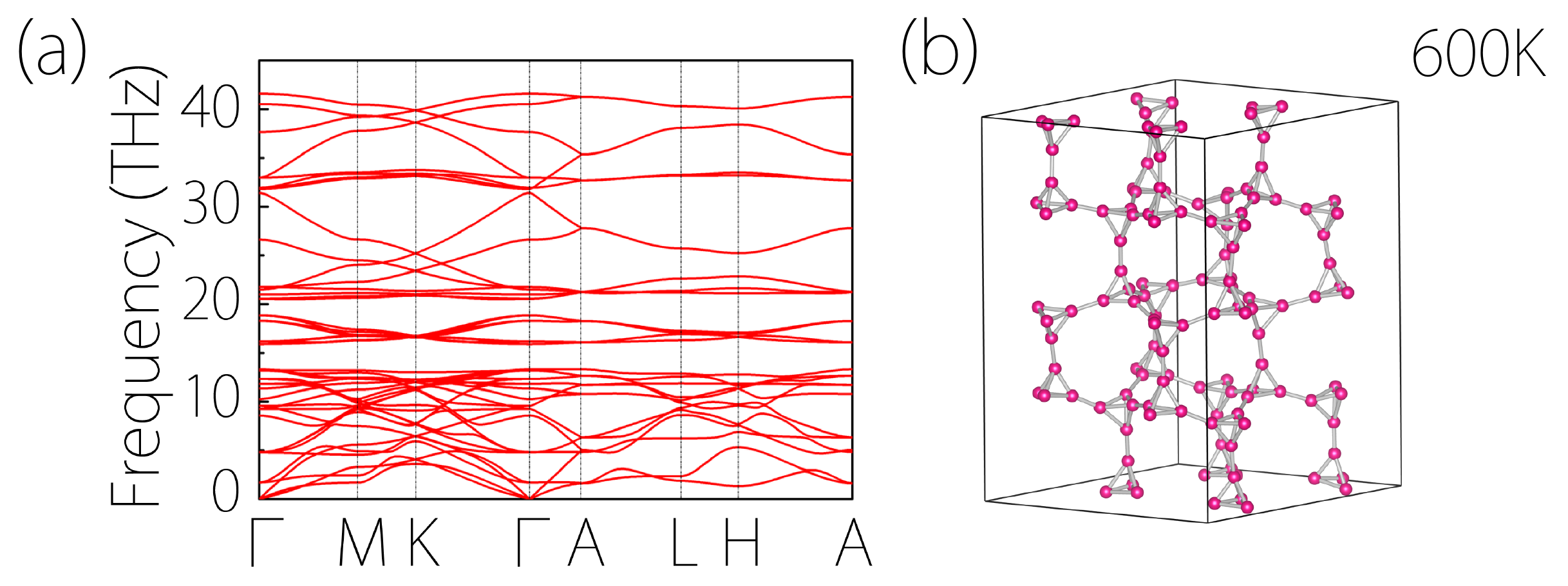}
	\caption{(a) Phonon dispersion of H-boron in the whole BZ. (b) Side view of the snapshot for the equilibrium structure of H-boron at the temperature of 600~K after 15~ps \emph{ab initio} molecular dynamics simulations.}
	\label{fig_phonon}
\end{figure}

The detailed structural parameters and basic properties of H-boron are listed in Table \ref{tab:I}. For comparison, we also list several other 3D boron allotropes. Note that for fair comparison, all data shown here are obtained by our own calculations using the same computational method. One notes that H-boron has similar properties as cF-B$_{8}$, which is expected because they are based on two closely related structures. Compared with other boron allotropes, H-boron has a smaller bulk modulus, consistent with its low density. Nevertheless, its bulk modulus ($\sim$70.42~GPa) is still larger than glass (35-55~GPa). In view of its very low mass density $\sim$0.91~g/cm$^{3}$, which is even less than water, the mechanical property of H-boron is in fact quite excellent. Consequently, H-boron may find potential applications as aerospace and cosmic materials.

In order to further examine the energetics, we calculated the total energies against the volume (equation of states, EOS) curves for seven typical boron allotropes, including H-boron, cF-B$_{8}$~\cite{37Getmanskii2017}, $\gamma$-B$_{28}$~\cite{14Gatti2009}, $\alpha$-Ga-type boron~\cite{13Haussermann2003}, $\alpha$-B$_{12}$~\cite{14Gatti2009}, $Pnnm$-B$_{16}$~\cite{34Yazyev2018}, and 3D-$\alpha'$ boron~\cite{33GaoYan2018} (Fig.~\ref{fig_energy}). Among them, $\alpha$-B$_{12}$ has the lowest energy, which is in good agreement with previous studies~\cite{14Gatti2009}. In comparison, the energy of H-boron, which is very close to cF-B$_{8}$, is at a local minimum, implying that this structure is energetically meta-stable. It is noted that with increasing crystal volume, H-boron becomes more and more energetically favorable. This suggests that H-boron may be more easily synthesized under an environment with negative pressure, which is similar to T-carbon~\cite{36Jinying2017}.

To examine whether H-boron is dynamically stable or not, we calculated the phonon spectrum for H-boron. As can be seen in Fig.~\ref{fig_phonon}(a), there is no imaginary phonon modes in the whole BZ, indicating that H-boron is dynamically stable. We also investigated the thermal stability of H-boron by carrying out \emph{ab initio} molecular dynamics simulations at finite temperatures. After heating H-boron to a targeted temperature of 600~K for 15~ps, we find that the structure maintains its integrity [see Fig.~\ref{fig_phonon}(b)], meaning that H-boron can be stable at 600~K.

\begin{figure}
	\centering
	\includegraphics[width=0.48\textwidth]{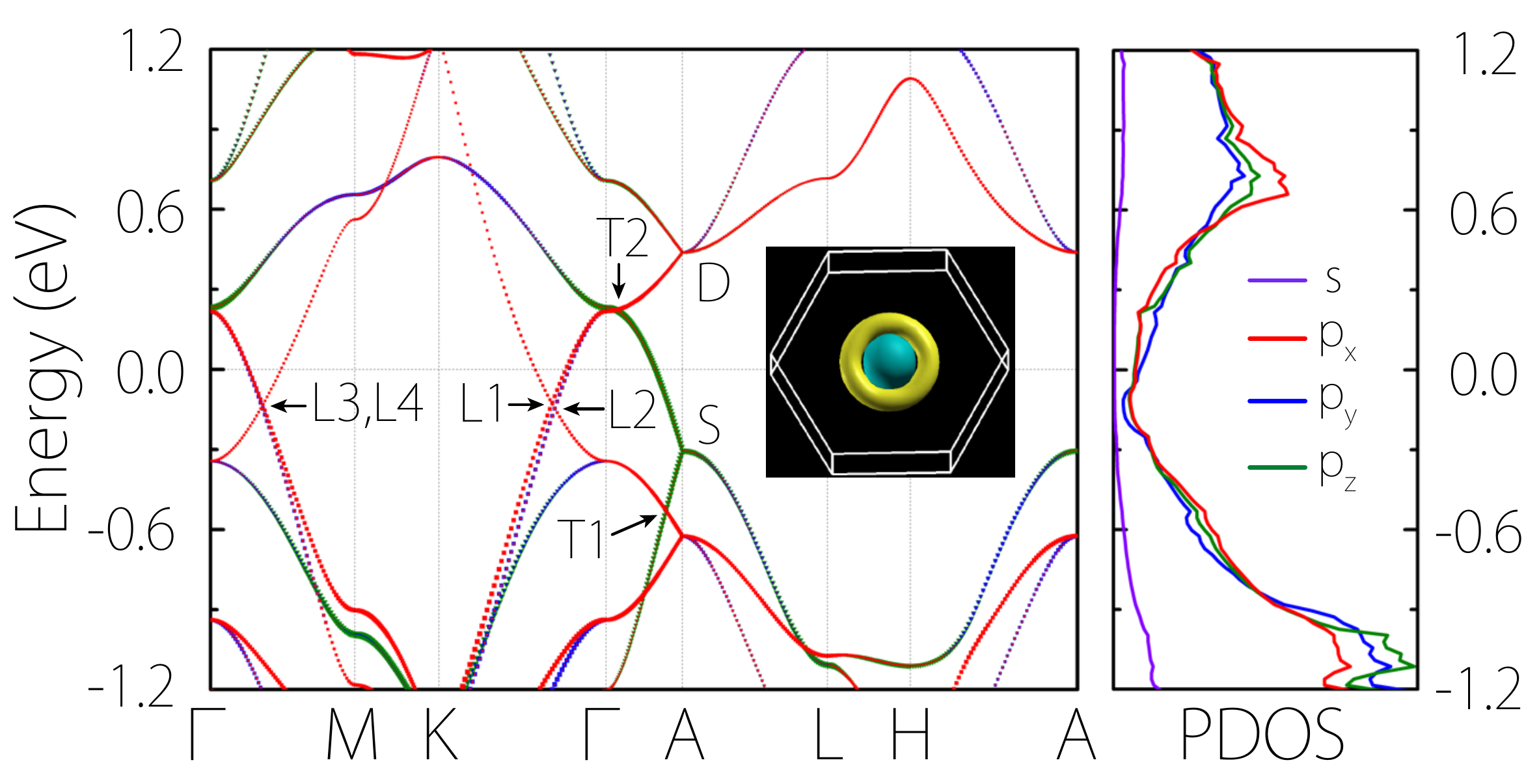}
	\caption{(a) Band structure of H-boron along high-symmetry paths in the BZ. Here, the size of the red, blue, and green dots indicates the weight of $p_{x}$, $p_{y}$ and $p_{z}$ orbitals of B atoms, respectively. The multiply degenerate nodal points are labeled as S, T1, T2, D along the $\Gamma$-$A$ path, L1 and L2 along the $K$-$\Gamma$ path, and L3 and L4 along the $\Gamma$-$M$ path. The top view of Fermi surface is shown in the inset. (b) Partial density of states (PDOS) for H-boron.}
	\label{fig_band}
\end{figure}

The most interesting property of H-boron lies in its electronic band structure, which we discuss in the following. Figure~\ref{fig_band}(a) shows the band structure of H-boron along the high-symmetry paths of the BZ. Note that the band structure is calculated without SOC, because SOC strength for boron is negligible. In the following discussion, the degeneracy is counted without considering spin. (If counting spin, the degeneracy appearing below would be doubled.) Obviously, H-boron is a metal without energy gap. From the projected density of states (PDOS), one can see that the low-energy states around the Fermi level are mainly from the boron $p$ orbitals. The density of states shows a dip around the Fermi level, exhibiting a semimetallic character.

\begin{figure*}
	\centering
	\includegraphics[width=0.8\textwidth]{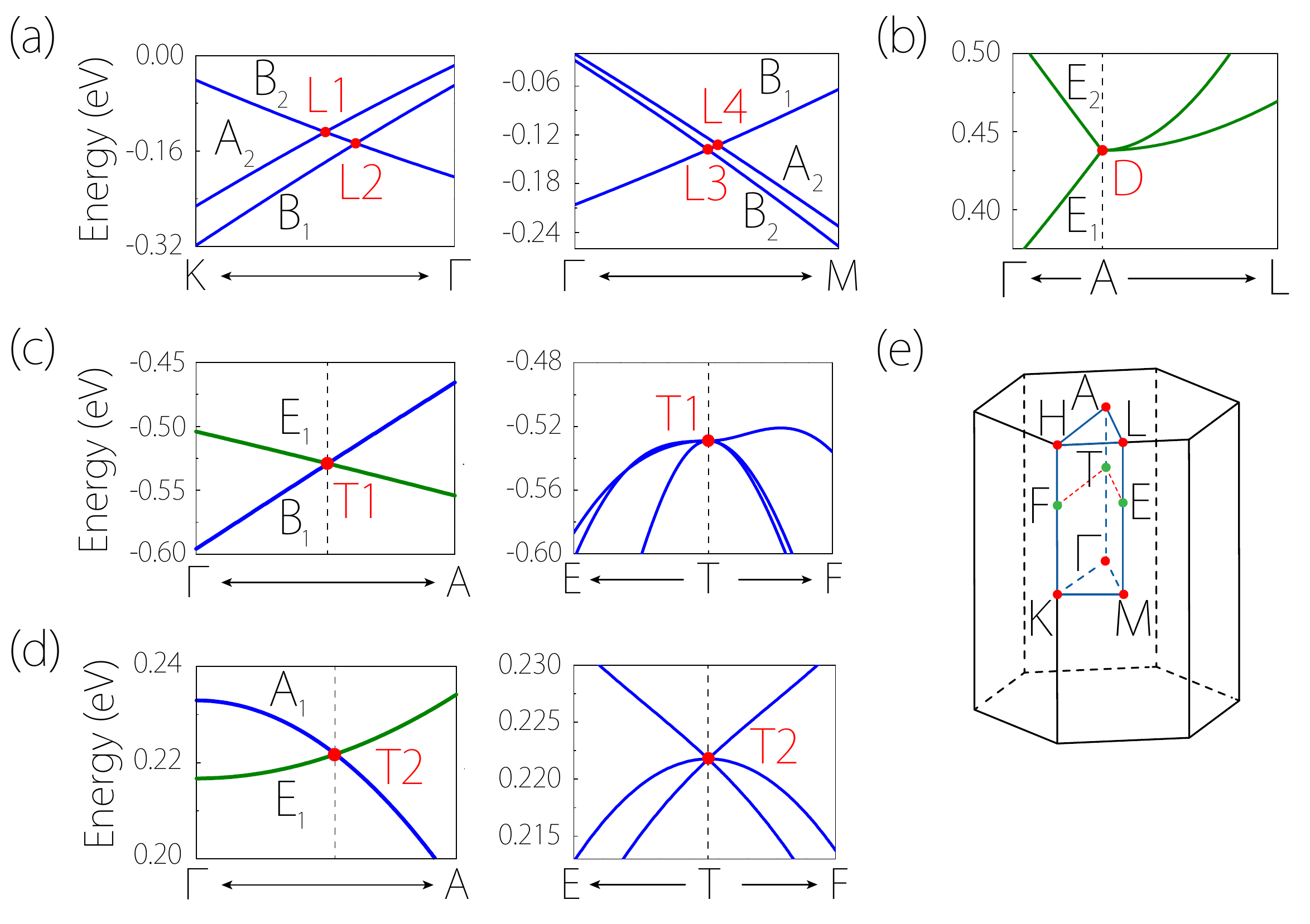}
	\caption{Enlarged band structures around the band degeneracies labeled in Fig.~\ref{fig_band}: (a) L1, L2 on $K$-$\Gamma$ (left panel) and L3, L4 on $\Gamma$-$M$ (right panel), (b) the fourfold degenerate point D, as well as the triply degenerate points (c) T1 and (d) T2. The right panels of (c) and (d) show the enlarged views of the respective band dispersions around T1 and T2 points along the $E$-$T$-$F$ path which is perpendicular to $\Gamma$-$A$. The irreducible representations along high-symmetry paths are also indicated. (e) Brillouin zone of H-boron. The point $T$ represents the triply degenerate point T1 or T2.}
	\label{fig_enlarge}
\end{figure*}

In Fig.~\ref{fig_band}(a), one observes that there are several band degeneracy points around the Fermi level. They can be classified into several groups. First, there are band crossing points $L1$ and $L2$ on $K$-$\Gamma$, as well as $L3$ and $L4$ on $\Gamma$-M [Fig.~\ref{fig_enlarge}(a)]. A careful scan of the BZ shows that these points are not isolated. They in fact lie on a network of nodal lines in the BZ [see Supplemental Material (SM) Fig. S1].

Second, there is a fourfold degenerate point D located at the time-reversal invariant momentum (TRIM) point $A$ on the BZ boundary (Fig.~\ref{fig_band}). Remarkably, we find that it belongs to a new type of spin-orbit-free band-degeneracy point: the semi-Dirac point. The band dispersion around the point is linear along $k_{z}$ ($\Gamma$-$A$ direction), while it is quadratic in the plane perpendicular to $k_{z}$ [Fig.~\ref{fig_enlarge}(b)]. To clarify its semi-Dirac character as well as to characterize the emergent fermions, we construct a $\bm{k} \cdot \bm{p}$ effective Hamiltonian around the point. At point $A$, we have the following key symmetry elements: six-fold screw rotation $\widetilde{\mathcal{C}}_{6z} = \{C_{6z}|00\frac{1}{2}\}$, inversion $\mathcal{P}$ and time reversal symmetry $\mathcal{T}$. At the $A$ point, the following algebra is satisfied:
\begin{equation}\label{eq_algebra}
\begin{aligned}
  \widetilde{\mathcal{C}}_{6z}^{6} &= e^{-i(3k_{z})} = -1, \\
  \widetilde{\mathcal{C}}_{6z}\mathcal{P} &= e^{-ik_{z}}\mathcal{P}\widetilde{\mathcal{C}}_{6z} = -\mathcal{P}\widetilde{\mathcal{C}}_{6z}.
\end{aligned}
\end{equation}
Since $[\widetilde{\mathcal{C}}_{6z},\mathcal{H}(\bm{k})] = 0$ at the $A$ point, one can use the eigenstates $|s_{z}\rangle$ of $\widetilde{\mathcal{C}}_{6z}$ with eigenvalues $s_{z} = e^{i\frac{2\pi}{6}(p+\frac{1}{2})}$, ($p=0,1,...,5$) as basis. Considering the anticommutation relation in Eq.~\ref{eq_algebra} and $\mathcal{T}$, we can obtain a set of states $\{|s_{z}\rangle,\mathcal{P}|s_{z}\rangle,\mathcal{T}|s_{z}\rangle,\mathcal{PT}|s_{z}\rangle\}$ with the eigenvalues $\{s_{z},-s_{z},s_{z}^{*},-s_{z}^{*}\}$. The crossing between four degenerate bands at $A$ requires $\{s_{z},-s_{z}\}\cap\{s_{z}^{*},-s_{z}^{*}\}=\varnothing$, which excludes the states with eigenvalues $\pm i$ ($p = 1,4$). Using the remaining states, one can have a set of orthogonal states $\{|e^{i\frac{\pi}{6}}\rangle,\mathcal{P}|e^{i\frac{\pi}{6}}\rangle,\mathcal{T}|e^{i\frac{\pi}{6}}\rangle,\mathcal{PT}|e^{i\frac{\pi}{6}}\rangle\}$ with eigenvalues $\{e^{i\frac{\pi}{6}},e^{i\frac{7\pi}{6}},e^{i\frac{11\pi}{6}},e^{i\frac{5\pi}{6}}\}$, forming a degenerate quartet at $A$. In the basis of the quartet, the symmetry operations have the following representation:
\begin{equation}\label{eq_representaion}
\begin{aligned}
  \widetilde{\mathcal{C}}_{6z} = e^{i\frac{\sigma_{z}}{2}\frac{2\pi}{6}} \otimes \sigma_{z}&, \qquad
  \mathcal{C}_{2x} = \sigma_{x} \otimes \sigma_{0}, \\
  \mathcal{P} = \sigma_{0} \otimes \sigma_{x}&, \qquad
  \mathcal{T} = \sigma_{x} \otimes \sigma_{0}\mathcal{K},
\end{aligned}
\end{equation}
where $\mathcal{K}$ is the complex conjugation operator, and $\sigma_{i}$ are the Pauli matrices. Constrained by these symmetries, the effective Hamiltonian takes the following form (to leading order in each momentum direction):
\begin{equation}\label{eq_hamit_D}
\begin{aligned}
\mathcal{H}_{D}(\bm{k}) = &\omega(\bm{k}) + \eta k_{z}\sigma_{z}\otimes\sigma_{z} \\
&+ [i\alpha k_{z}k_{-}\sigma_{+}\otimes\sigma_{0} + \beta k_{+}^{2}\sigma_{+}\otimes\sigma_{x} + H.c.],
\end{aligned}
\end{equation}
where $\omega(\bm{k}) = \epsilon_{0} + \epsilon_{1}(k_{x}^2+k_{y}^{2}) + \epsilon_{2}k_{z}^{2}$, $k_{\pm}=k_{x}\pm ik_{y}$, $\sigma_{\pm}=\sigma_{x}\pm i\sigma_{y}$, $\bm{k}$ is measured from the Dirac point, and $\alpha$, $\beta$, $\eta$ and $\epsilon_{i=0,1,2}$ are real parameters whose value can be determined by fitting the DFT band structure. The model clearly shows that the dispersion is linear along $k_{z}$, and quadratic in the $k_{x}-k_{y}$ plane, hence may be termed as a 3D semi-Dirac point. Previously, the concept of semi-Dirac point was discussed in 2D systems~\cite{46Hasegawa2006,47Victor2009}, where the dispersion is linear along one direction and quadratic along the other direction. To our knowledge, this is the first time that a 3D semi-Dirac point is revealed.

Third, point T1 on the $\Gamma$-$A$ path is a triply degenerate point. It is formed by the intersection between a doubly degenerate band and a nondegenerate band [Fig.~\ref{fig_enlarge}(c)]. The crossing bands correspond to the $B_{1}$ and $E_{1}$ irreducible representations of the $C_{6v}$ symmetry group. Using these states as basis, we can obtain the effective Hamiltonian around $T1$, given by
\begin{equation}\label{eq_hamit_QT}
\begin{aligned}
&\mathcal{H}_{QT}(\bm{k}) = \omega_{0}(\bm{q}) + \\
&\left[
\begin{matrix}
A_{0}q_{z} + A_{1}(q_{x}^{2}+q_{y}^{2}) + A_{2}q_{z}^{2} & -2iBq_{x}q_{y} & iB(q_{x}^{2}-q_{y}^{2}) \\ 	
2iBq_{x}q_{y} & C(q_{x}^{2}-q_{y}^{2}) & -2Cq_{x}q_{y} \\
-iB(q_{x}^{2}-q_{y}^{2}) & -2Cq_{x}q_{y} & -C(q_{x}^{2}-q_{y}^{2}) \\
\end{matrix}
\right],
\end{aligned}
\end{equation}
where $\omega(\bm{k}) = \epsilon_{0} + \epsilon_{1}q_{z} \epsilon_{2}(q_{x}^2+q_{y}^{2}) + \epsilon_{3}q_{z}^{2}$, $\bm{k}$ is measured from $T1$, $A_{0,1,2}$, $B$, $C$ and $\epsilon_{0,1,2,3}$ are real parameters whose value can be determined by fitting the DFT band structure. Interestingly, one notes that the dispersion around the triply degenerate point is of quadratic type in the horizontal plane while linear in the $k_z$ direction [Fig.~\ref{fig_enlarge}(c)]. Previous works have reported triple-point fermions in several materials but the dispersion around them exhibits linear features~\cite{30Zhongcy2017,48DaiXi2016,49Ziming2016,50ZhongDai2016,51Bradlyn2016,52XiaomingYu2017}. In contrast, the triply degenerate fermions here have a higher order (quadratic) in-plane energy dispersion, which is of a distinct new type and may be termed as quadratic triple point.

Fourth, point T2 is also a triply degenerate point but the dispersion around them is of linear type [Fig.~\ref{fig_enlarge}(d)]. Thus, they share similar features as those reported in the centrosymmetric spin-orbit-free materials h-C28~\cite{53Chengyong2019}. We find that it is formed by the crossing between the $A_{1}$ band and the $E_{1}$ band of the $C_{6v}$ group along the $\Gamma$-$A$ path. Using the $A_{1}$ and $E_{1}$ states as basis, we can obtain the effective Hamiltonian (up to linear order) around $T2$, given by
\begin{equation}\label{eq_hamit_LT}
\begin{aligned}
\mathcal{H}_{LT} &= \omega_{0}(\bm{q}) + \left[
\begin{matrix}
Aq_{z} & Bq_{y} & Bq_{x} \\ 	
Bq_{y} & 0 & 0 \\
Bq_{x} & 0 & 0 \\
\end{matrix}
\right],
\end{aligned}
\end{equation}
where $\omega(\bm{k}) = \epsilon_{0} + \epsilon_{1}q_{z}$, $A$, $B$, $C$ and $\epsilon_{0,1}$ are real parameters.

Finally, we examine the degenerate low-energy bands (including the $S$ point) in the $k_{z}=\pi$ plane (Fig.~\ref{fig_enlarge}), which are actually parts of a nodal surface formed by the linear crossing between the two bands along the $k_{z}$ direction. The presence of the nodal surface is dictated by the combination of a twofold screw rotation $\mathcal{S}_{2z} = \{C_{2z}|00\frac{1}{2}\}$ and $\mathcal{T}$. In the absence of SOC, since $[\mathcal{T}, \mathcal{S}_{2z}]=0$, $(\mathcal{T}\mathcal{S}_{2z})^{2}=e^{-ik_{z}}$. One notes that the $k_{z}=\pi$ plane is invariant under $\mathcal{T}\mathcal{S}_{2z}$, and the antiunitary symmetry satisfies $(\mathcal{T}\mathcal{S}_{2z})^{2}=-1$ which indicates a Kramers-like degeneracy on the whole plane. Away from $k_{z}=\pi$ plane, the degeneracy is lifted due to the loss of symmetry protection. Consequently, the band crossing at $k_{z}=\pi$ plane forms a nodal surface [SM Fig. S1]. A similar argument has been presented in the previous discussion of spin-orbit-free materials~\cite{32Weikang2018,54QiFeng2016,55WeikangZhang2018}.

\section{Discussion and summary}\label{sec_discussion}
We have predicted a new 3D boron allotrope, the H-boron. The material has not been realized yet. However, in view of its good stability and the recent success in the synthesis of T-carbon (which is closely related to H-boron), we expect that the material can be realized in near future. Here we also try to propose a possible scheme for its synthesis. One may start from the minimum basic unit of graphane. The carbon atoms can be replaced with boron tetrahedrons as in cF-B$_{8}$~\cite{37Getmanskii2017}. Then by stacking the single layers of borane, one can obtain the 3D H-boron with multilayered structure after a dehydrogenation reaction. These steps are schematically illustrated in the SM Fig. S2.

To facilitate the detection of H-boron, we have simulated the X-ray diffraction (XRD) pattern of H-boron along with several other boron allotropes (Fig. S3 in the SM), which can be used as structural indicators in future experiments.

We have shown that H-boron is a topological metal with multiple types of spin-orbit-free emergent fermions. Particularly, the 3D semi-Dirac fermion and the quadratic triple-point fermion are observed here for the first time. In experiment, these novel emergent fermions can be probed by the angle-resolved photoemission spectroscopy (ARPES). Especially, the band degeneracies below the Fermi level should be readily detected by ARPES.

Finally, we note that the presence of large cavities in H-boron may bring several advantages. First, this leads to the low density of H-boron, which we have discussed above. Second, this makes it easier to adjust the electronic property of H-boron by stress or pressure [see SM Figs. S4 and S5]. In addition, the cavities may accommodate hydrogen or Li atoms for energy storage applications.

In summary, we have predicted a novel 3D boron allotrope, namely H-boron, by using first-principles electronic structure calculations. We find that the material has good stability and excellent mechanical property. Its very low mass density may permit potential applications as aerospace and cosmic materials. Importantly, we reveal that H-boron is a new topological metal with multiple types of spin-orbit-free emergent fermions. In particular, the 3D semi-Dirac fermion and the quadratic triple-point fermion found here have not been discussed before. Thus, our work not only discovers a new boron allotropic material, it also reveals a platform to explore the fascinating physics associated with new emergent fermions.

\begin{acknowledgments}
We wish to thank D. L. Deng for helpful discussions. This work was supported by the National Key R\&D Program of China (Grant No. 2017YFA0302903), the National Natural Science Foundation of China (Grants No. 11774422, No. 11774424, and No. 11804039), and the Singapore Ministry of Education AcRF Tier 2 (MOE2015-T2-2-144). Computational resources were provided by the Physical Laboratory of High Performance Computing at Renmin University of China.
\end{acknowledgments}

\bibliography{H-boron-new.bib}

\end{document}